\documentclass[aps,pra,twocolumn,superscriptaddress,longbibliography,nobalancelastpage,floatfix]{revtex4-2}

\usepackage{graphicx} 
\usepackage[hypertexnames=false]{hyperref}
\usepackage{physics}
\usepackage{orcidlink}

\begin{document}
\title{Process tensor approaches to modeling two-dimensional spectroscopy}
\author{Roosmarijn de Wit}
\affiliation{SUPA, School of Physics and Astronomy, University of St Andrews, St Andrews, KY16 9SS, United Kingdom}
\author{Jonathan Keeling}
\affiliation{SUPA, School of Physics and Astronomy, University of St Andrews, St Andrews, KY16 9SS, United Kingdom}
\author{Brendon W. Lovett}
\affiliation{SUPA, School of Physics and Astronomy, University of St Andrews, St Andrews, KY16 9SS, United Kingdom}
\author{Alex W. Chin}
\affiliation{Sorbonne Université, CNRS, Institut des NanoSciences de Paris, 4 place Jussieu, 75005 Paris, France}
\date{\today}

\begin{abstract}
    Problems in the field of open quantum systems often involve an environment that strongly influences the dynamics of excited states. Here we present a numerical method to model optical spectra of non-Markovian open quantum systems. The method employs a process tensor framework to efficiently compute multi-time correlations in a numerically exact way. To demonstrate the efficacy of our method, we compare 2D electronic spectroscopy simulations produced through our method to Markovian master equation simulations in three different system-bath coupling regimes.
\end{abstract}

\maketitle

\section{Introduction}
Two-dimensional electronic spectroscopy (2DES) is a valuable tool for probing photophysical processes in light-harvesting systems~\cite{mukamel:1995, cho:2008, collini:2021, engel:2007, lewis:2012, yeh:2018, wang:2019, bukarte:2020, cao:2020, kim:2021, higgins:2021}. Generally, these systems do not exist in isolation, and measured quantities are significantly impacted by a vibrational environment. 2DES has yielded key insights into the role of such environment-mediated processes ~\cite{lewis:2012, yeh:2018, wang:2019, bukarte:2020, cao:2020, higgins:2021}. However, extracting the underlying mechanism remains challenging. Theoretically, realistic environments are often non-Markovian, meaning that time-local (Markovian) equations of motion are not sufficient~\cite{breuer:2002, vega:2000}. Experimentally, the complexity of measured signals can make it difficult to accurately interpret observed spectral features. For example, the as-yet-unclear origin of long-lived coherences in biological light absorbing pigments highlights the necessity of accurate theoretical models to explain experimental observations~\cite{cao:2020,chin:2013, lim:2015, yeh:2018, wang:2019, bukarte:2020, higgins:2021, caycedo:2022}. 

Here we present an efficient tensor network method to simulate optical spectra of non-Markovian open quantum systems. To demonstrate its capabilities, we introduce a model describing the essential features of a broad class of molecular chromophores. It consists of a three-level electronic system coupled via a vibrational bath, enabling the intramolecular transfer of energy between system states. We first compare numerical results to master equations in weak and strong (polaronic) system-bath coupling regimes to test the validity of our method. Then, we  highlight key differences between the 2D spectra obtained via these methods in an intermediate coupling regime. 
Namely, we find that there are significant differences in peak positions at high temperatures, related to the Lamb shift. At low temperatures, we furthermore find that a master equation underestimates the dephasing time of the optical response, as has been observed in previous work~\cite{chin:2013}.

Mathematically, a 2DES signal is expressed as a sum of four-time correlation functions that encode all possible light-matter interactions. A common approach to calculating multi-time correlation functions is to use the quantum regression theorem, which is based on the Born-Markov approximation~\cite{breuer:2002}. In many realistic scenarios however, a Markovian picture is not sufficient. Even if it accurately represents the reduced dynamics of the system, a Markovian description does not necessarily capture multi-time correlations correctly~\cite{guarnieri:2014}. Some light-harvesting complexes are furthermore known to operate in the intermediate coupling regime, where electronic couplings within the system are comparable in strength to system-bath interactions and a Markovian approximation is not justified.~\cite{ishizaki:2010, fassioli:2014, jumper:2018}. 

Simulating non-Markovian open quantum systems is a challenging task. Nonetheless, numerous numerical methods have been developed to achieve this~\cite{vega:2000, orus:2014, pollock:2018, jorgensen:2019, fux:2021, cygorek:2022, ng:2023, thoenniss:2023, thoenniss2:2023, cygorek:2023update, link:2023, yeh:2018, tanimura:1988, tanimura:2006, tanimura:2020, prior:2010, chin:2010, chin:2013, tamascelli:2019, somoza:2019, mascherpa:2020, caycedo:2022, sirkina:2023, oqupaper}. Most relevant to this work, one group of such methods uses a process tensor (PT) formalism to capture the influence of the environment on the system. Importantly, the process tensor is cast into a matrix product operator format (PT-MPO), such that only the physically most relevant part of the Hilbert space is efficiently represented \cite{orus:2014, jorgensen:2019, pollock:2018, fux:2021, cygorek:2022, ng:2023, thoenniss:2023, thoenniss2:2023, cygorek:2023update, link:2023, oqupaper}. Moreover, since PT-MPO methods rely on constructing a reduced density matrix description of the system, they are particularly well-tailored to calculating temporal correlations.

The remainder of this paper is organized as follows.  Section~\ref{sec:model} introduces the model, Section~\ref{sec:2DES} specifies the multi-time correlation functions relevant to our results, and Section~\ref{sec:PTMPO} provides an overview of the PT-MPO computation method. The results are presented in Sections~\ref{sec:MECompare} and~\ref{sec:Differences}; Section~\ref{sec:MECompare} compares spectra obtained via PT-MPO methods versus Markovian master equations in weak and strong system-bath coupling regimes, while Section~\ref{sec:Differences} further analyses the PT-MPO results in an intermediate coupling regime.

\section{The model}
\label{sec:model}
\begin{figure}
\includegraphics[width=8cm]{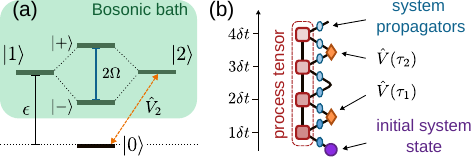}
\caption{(a) Sketch of our model, consisting of a three-level system coupled to a bosonic bath. (b) Tensor network of four time steps for calculating a two-time correlation function with a PT-MPO.}  
\label{fig1}
\end{figure}

As illustrated in Fig.~\ref{fig1}(a), our model consists of a three-level system with excited states linearly coupled to a bosonic bath, $\hat{H}=\hat{H}_S+\hat{H}_B+\hat{H}_I$ with:
\begin{align}
    \label{ham} 
    \hat{H}_S &= (\epsilon + \lambda)\pqty{ \dyad{1}{1} + \dyad{2}{2}} + \Omega\pqty{\dyad{1}{2} + \mathrm{H.c.}},\\
    \hat{H}_B &= \sum_k\,\omega_k b_k^{\dagger}b_k
    , 
    \\  \hat{H}_I&=
    \pqty{\dyad{1}{1} - \dyad{2}{2}}\sum_k(g_k b_k + g_k^{*}b_k^{\dagger}).
\end{align}
Here, the three terms correspond to the system ($\hat{H}_S$), bath ($\hat{H}_B$) and system-bath interaction ($\hat{H}_I$). The energy of the excited states $\pqty{\ket{1}, \ket{2}}$ is given by the bare energy $\epsilon$ plus the reorganization energy of the bath $\lambda$ (defined below), and electronic coupling $\Omega$. Physically, $\ket{1}$ and $\ket{2}$ could describe two excited states of a chromophore with ground state $\ket{0}$. These excited states could be intramolecular in nature, as in the $S_1, S_2$ states found in many biological pigments, organic dye molecules and nanoparticles  \cite{oviedo2010dynamical,dunnett2021influence,dufour2017engineering}, or the lowest single-particle excitations of coupled chromophores, such as excitonically coupled H-dimers, or charge-transfer complexes \cite{wang2021twisted,rafiq2021interplay}. For clarity in later discussions, we imagine our system to be an excitonic dimer so that the coupling $\Omega$ can be thought of as creating electronic eigenstates ({\it i.e.} excitons; $\ket{\pm} = \frac{1}{\sqrt{2}} \pqty{\ket{1} \pm \ket{2} }$) that  are delocalized across the internal chromophore monomers with energies $E_{\pm} = \epsilon + \lambda \pm \Omega$ (Fig.~\ref{fig1}(a)). Since chromophores in biological systems often display photophysics (dynamics)  on a picosecond time scale, we set $\epsilon=5$~ps$^{-1}$ and $\Omega = 0.2-2$~ps$^{-1}$~\cite{wang:2019}. The bath is described by vibrational modes with frequency $\omega_k$ and creation (annihilation) operators $b_k^\dagger$ ($b_k$). The system-bath coupling constants $g_k$ are characterized by the spectral density $J(\omega)~=~\sum_k\abs{g_k}^2\delta (\omega - \omega_k)$. We set $J(\omega)$ to be Ohmic, as is common in the modeling and interpretation of molecular spectroscopies~\cite{mukamel:1995, cho:2008}:
\begin{equation}
    J(\omega) = 2 \alpha \omega e^{-\frac{\omega}{\omega_c}},
    \label{sd}
\end{equation}
where $\alpha$ is a dimensionless parameter for the system-bath coupling strength and $\omega_c=3.04$~ps$^{-1}$ represents the bath cut-off frequency. The reorganization energy is then given by $\lambda = \int_0 ^\infty \dd\omega \frac{1}{\omega} J(\omega) = 2\alpha\omega_c$. We set the initial state $\rho_0$ to be a product state of the system ground state $\ket{0}$ and the thermal state of the environment at temperature $T$.

\section{2D Electronic Spectroscopy}
\label{sec:2DES}

2DES involves probing a quantum system with three short laser pulses while systematically varying the time delays between each pulse. For a comprehensive review on 2DES, we refer to Refs.~\cite{cho:2008,collini:2021}. Here, we will consider the measured third-order optical response, which consists of a series of four-time correlation functions. In order to predict the spectra resulting from this model, we will initially simulate laser pulses that induce transitions between the ground state and excited state $\ket{2}$, corresponding to the dipole operator $\hat{V}_2=\ketbra{0}{2}+\mathrm{H.c.}$. We will assume  we are in the semi-impulsive limit, such that we can neglect the temporal width of the laser pulses~\cite{mukamel:1995, cho:2008, hamm:2011}. We will furthermore make the rotating-wave approximation and assume that the laser pulses applied at times $\tau_i$ are time-ordered~\cite{mukamel:1995, hamm:2011}. Consequently, only the following four multi-time correlation functions contribute to the 2D signal:
\begin{flalign}
    R_1 &= \Tr \big[ \hat{V}_2(\tau_4)\hat{V}_2(\tau_1)\rho_0\hat{V}_2(\tau_2)\hat{V}_2(\tau_3) \big] \nonumber \\
    R_2 &= \Tr \big[\hat{V}_2(\tau_4)\hat{V}_2(\tau_2)\rho_0 \hat{V}_2(\tau_1)\hat{V}_2(\tau_3) \big] \nonumber \\
    R_3 &= \Tr \big[\hat{V}_2(\tau_4)\hat{V}_2(\tau_3)\rho_0 \hat{V}_2(\tau_1)\hat{V}_2(\tau_2) \big] \nonumber \\
    R_4 &= \Tr \big[\hat{V}_2(\tau_4)\hat{V}_2(\tau_3) \hat{V}_2(\tau_2)\hat{V}_2(\tau_1) \rho_0 \big],  
    \label{R}
\end{flalign}
which are commonly grouped together as rephasing ($R_2, R_3$) and non-rephasing ($R_1, R_4$) pathways, according to the experimental phase matching conditions~\cite{collini:2021, mukamel:1995, cho:2008}. The real parts of these pathways in the frequency domain sum up to give the total 2D spectrum; this is a function of an excitation ($\omega_{exc}$) and detection frequency ($\omega_{detec}$) corresponding to Fourier transforms of the first ($\tau_2 - \tau_1$) and third time delay ($\tau_4 - \tau_3$) between pulses. Note that in this work, we will only consider 2D spectra for which the so-called waiting time $\tau_3-\tau_2$ is zero. For the linear response, there is only a single pathway: $\Tr \big[ \hat{V}_2(\tau_2)\hat{V}_2(\tau_1)\rho_0\big] $, which is similarly Fourier transformed with respect to $\tau_2-\tau_1$ to obtain a linear absorption spectrum as a function of frequency $\omega$.

\section{System response functions and PT-MPOs}
\label{sec:PTMPO}

In order to calculate the multi-time correlation functions in Eq.~\eqref{R}, we will employ the Time Evolving MPO (TEMPO) method to acquire a PT-MPO that captures any possible non-Markovian effects on the system dynamics~\cite{strathearn:2017, pollock:2018,  strathearn:2018, fux:2021, fux:2023, oqupaper}. The Python code for PT-TEMPO is available in the open-source package OQuPy~\cite{oqupy}. The following sections will give further details on how multi-time correlation functions are calculated in a PT framework, and how a PT-MPO is constructed with the PT-TEMPO method. For a more detailed description of the tensor network methods used in this work, we refer the reader to Refs. \cite{oqupaper} (overview of the OQuPy package), \cite{strathearn:2018} (general TEMPO framework), \cite{jorgensen:2019, fux:2021} (PT-TEMPO) and \cite{fux:2023, oqupaper} (calculating multi-time correlations with PT-MPOs).

\subsection{Calculating multi-time correlation functions}
Let us consider time-ordered multi-time correlations of a general form:
\begin{equation}
    R = \expval{\prod_{p=1}^{P} \hat{V}_p (\tau_p)} =  \expval{\prod_{p=1}^{P} \hat{V}_p (M_p \delta t)}
    \label{C}
\end{equation}
such that $P$ operators ${\hat{V}_p}$ are applied at times $\tau_p = M_p \delta t$, where $\delta t$ denotes the time step. 
A formal approach to calculating $R$ is to consider the time evolution of the total ({\it i.e.} system and bath) density matrix $\rho_0$. First, $\rho_0$ is propagated up to $M_1$ time steps according to the formal solution of the von Neumann equation:
\begin{equation}
    \rho(M_1 \delta t) = e^{\mathcal{L} M_1 \delta t} \rho_0,
\end{equation}
with the total Liouvillian $\mathcal{L} = -i\bqty{\hat{H},\cdot}$.
Then, after application of the first operator $\hat{V}_1$, the resulting density matrix is propagated for $M_2 - M_1$ time steps, and so on, up to the application of the final operator $\hat{V}_P$. Thus, we can write Eq.~\eqref{C} in terms of the time propagators:
\begin{equation}
    R =  \Tr \bqty{\prod_{p=1}^{P} \pqty{\hat{V}_p^{L,R} \bqty{e^{\mathcal{L} \delta t (M_p - M_{p-1})}      }   }\rho_0},
    \label{C2}
\end{equation}
where $M_0 = 0$. Here the superscript $L,R$ signifies that the super-operator $\hat{V}$ can act to the left ($\hat{V}^L = \hat{V}\cdot$) or to the right ($\hat{V}^R = \cdot \hat{V}$) of the density matrix.

However, applying this approach directly would require direct evolution of the full density matrix $\rho_0$, which is not practical. Instead, we calculate Eq.~\eqref{C2} by constructing a tensor network in which the influence of the bath on the system is encoded in a PT-MPO. The PT is a multi-linear map from operations performed on a system, at a sequence of time steps, to its final state. Crucially, since the PT is completely independent of the system Hamiltonian $\hat{H}_S$, it can be calculated before specifying any system parameters, the initial system state $\rho^S_0$ or the operators $\hat{V}$ in Eq.~\eqref{C2}. To separate system and bath propagation, we perform a second-order Suzuki-Trotter splitting~\cite{suzuki:1992}:
\begin{equation}
    \bqty{e^{\mathcal{L}\delta t}}^M = \bqty{ e^{\mathcal{L}_S \frac{\delta t}{2}}  e^{\mathcal{L}_E \delta t } e^{\mathcal{L}_S \frac{\delta t}{2} }}^M + \mathcal{O}\pqty{\delta t ^3},
    \label{trotter2}
\end{equation}
such that $\mathcal{L}_S = -i\bqty{\hat{H_S},\cdot}$ and $\mathcal{L}_E = -i\bqty{\hat{H}_B + \hat{H}_I,\cdot}$. Once the PT-MPO is constructed (see the next section), multi-time correlation functions are calculated by combining the PT-MPO with the operators $\hat{V}_p$ at times $M_p \delta t$ and system propagators $K =  e^{\mathcal{L}_S \frac{\delta t}{2}}$.

Fig.~\ref{fig1}(b) shows a tensor network diagram of a PT-MPO for four time steps (red squares) and illustrates how multi-time correlations are calculated. The tensor network is expressed in Liouville space, such that density matrices are represented by vectors and superoperators by matrices. 
Diagrammatically, the initial system state $\bqty{\rho^S_0}^j$ therefore corresponds to a tensor with a single leg (purple circle), where the index $j$ runs from 1 to the squared dimension of the system Hilbert space. Similarly, the dipole operators $\bqty{V}_j^{j^\prime}$ (orange diamonds) and system propagators $\bqty{K}_j^{j^\prime}$ (two per single time step, blue ovals) have two legs. Thus, multi-time correlation functions are computed by applying the system propagators and dipole operators as a set of interventions at the relevant time steps and subsequently tracing over the bonds in the network to obtain $R$~\cite{gribben:2022, fux:2023, oqupaper}. Since the final dipole operator $\hat{V}_P$ is applied at the end of the chain, we only have to run the simulation once to calculate $R$ over a range of the final times $M_P \delta t$. When varying some earlier time arguments $M_p \delta t$, {\it e.g.} over a range of $i$ time steps, the simulation is repeated $i$ times, moving the position of $\hat{V}_p$ in the tensor network (Fig.~\ref{fig1}(b)) with each repetition. For further details on the required computational resources, see Appendix A and Ref.~\cite{oqupaper}. 

\subsection{PT-MPO construction}

The method described above is general, and there exist multiple algorithms to construct the PT-MPO \cite{orus:2014, jorgensen:2019, pollock:2018, fux:2021, cygorek:2022, ng:2023, thoenniss:2023, thoenniss2:2023, cygorek:2023update, link:2023, oqupaper}. In this work, we employ the PT-TEMPO method to construct a PT-MPO, which we will summarize here~\cite{jorgensen:2019, fux:2021, oqupy, oqupaper}. For readers already familiar with PT-MPOs, this section can be skipped.

The general TEMPO framework expresses the (non-Markovian) impact of the bath on the system in terms of a discretized Feynman-Vernon influence functional~\cite{strathearn:2018, feynman:1963, makri:1995, makri2:1995}. In PT-TEMPO, the parts of the TEMPO network corresponding to the influence functional are contracted into a PT-MPO that does not depend on the system propagators. To illustrate how the tensor network is constructed, let us consider the propagation of the system density matrix $\rho^S$ for $M$ time steps, such that the total propagation consists of $M$ propagators over short time steps $\delta t$: $\bqty{e^{\mathcal{L}\delta t}}^M$. We start with the Suzuki-Trotter splitting in Eq.~\eqref{trotter2}.
Then, resolutions of identity are added between each system and bath propagator. By tracing over the environment, we obtain a path sum over system states composed of the discretized influence functional and system propagators:
\begin{multline}
    \rho^S_{j_M}(M \delta t) = \sum_{\substack{j_0, \ldots, j_{M-1} \\ j^\prime_0, \ldots, j^\prime_{M-1}}} \Bigg[\prod_{m=0}^{M-1} K_m (j_{m+1}, j^\prime_m) \\
   \cp  \pqty{\prod_{k=0}^{m} I_k (j^\prime_m, j^\prime_{m-k})} K_m (j^\prime_m, j_{m})\Bigg] \rho^S_{j_0}.
   \label{prop}
\end{multline}
In this expression, the system propagators are given by $K (j, j^\prime) =  \bqty{\exp \pqty{\mathcal{L}_S      \frac{\delta t}{2}}       }_{j,j^\prime}$. Because each propagator evolves $\rho_j^S$ by half a time step, we require two indices: $j_m$ runs over each full step, while $j_m^\prime$ connects the pairs of propagators within a step. The influence functions $I_k(j,j^\prime)$ capture the effect of the bath on the system, and connect system states separated by $k$ time steps. For the exact form of $I_k$, see Ref.~\cite{strathearn:2018}. 

\begin{figure}
\includegraphics[width=8cm]{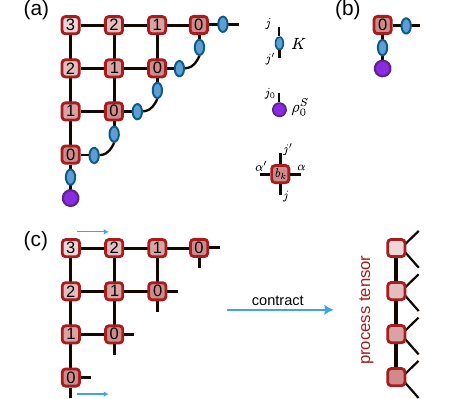}
\caption{(a) The TEMPO tensor network. The bath tensors $b_k$ are represented by red squares, labeled with index $k$. (b) The network for a single time step. (c) The bath tensors of the TEMPO network are horizontally contracted into a process tensor.}
\label{fig2}
\end{figure}

Fig.~\ref{fig2}(a) illustrates how Eq.~\eqref{prop} can be constructed as a tensor network. Per time step, we apply two system propagators (blue ovals, each evolving by half a time step), represented by the rank-2 tensors $K_j^{j^\prime}$. The influence functions are incorporated as the following bath tensors:
\begin{equation}
    \bqty{b_k}^{\alpha,j}_{\alpha^\prime, j^\prime} = \delta^{j}_{j^\prime}\delta^{\alpha}_{\alpha^\prime} I_k(\alpha,j),
    \label{bathtensor}
\end{equation}
which are drawn as red squares (labeled with index $k$) in Fig.~\ref{fig2}(a). At the left and top edges of the network, we instead require tensors that lack the $\alpha^\prime$ or $j^\prime$ leg. These are obtained from Eq.~\eqref{bathtensor} by respectively omitting the $\delta^\alpha_{\alpha^\prime}$ or $\delta^j_{j^\prime}$ Kronecker deltas. 
To illustrate how Eq.~\eqref{prop} maps to a tensor network, consider the propagation of $\rho^S$ for a single time step ($M=1$):
\begin{equation}
    \rho^S_{j_1}= \sum_{j_0,j^\prime_0}K_0(j_1, j^\prime_0)I_0(j^\prime_0,j^\prime_0)K_0(j_0^\prime, j_0)\,\rho^S_0,
\end{equation}
which can be written as the tensor contraction
\begin{equation}
    \bqty{\rho^S}^{j_1} = \sum_{j_0,j^\prime_0,\alpha}  \, \bqty{K_0}_{\alpha}^{j_1}[b_0]^\alpha_{j^\prime_0} \,\bqty{K_0}_{j_0}^{j^\prime_0}\,\bqty{\rho^S}^{j_0},
\end{equation}
shown in Fig.~\ref{fig2}(b). 

The number of bath tensors---and thus the bath memory---builds up with each time step, such that at time $m \delta t$, $m$ bath tensors are added to the tensor network (Fig.~\ref{fig2}(a)). We can introduce a maximal memory time of the bath, set by the parameter $\Delta K_{max}$, which ensures that the number of bath tensors added to the network per time step stops increasing after $\Delta K_{max}$ time steps. 

With PT-TEMPO, the bath tensors of the TEMPO network shown in Fig.~\ref{fig2}(c) are contracted into a PT-MPO using a sequence of standard tensor contraction and compression techniques~\cite{orus:2014}. Through the truncation of singular values, the compression steps ensure that the tensor bond dimensions (and thus their size) are kept to a minimum. The truncation threshold is determined by the convergence parameter $\epsilon_{rel}$, which sets the upper bound to the singular values to be discarded, relative to the largest singular value in the tensor. For further information on the convergence parameters used in this work, see Appendix B.

\begin{figure*}
\includegraphics[width=16cm]{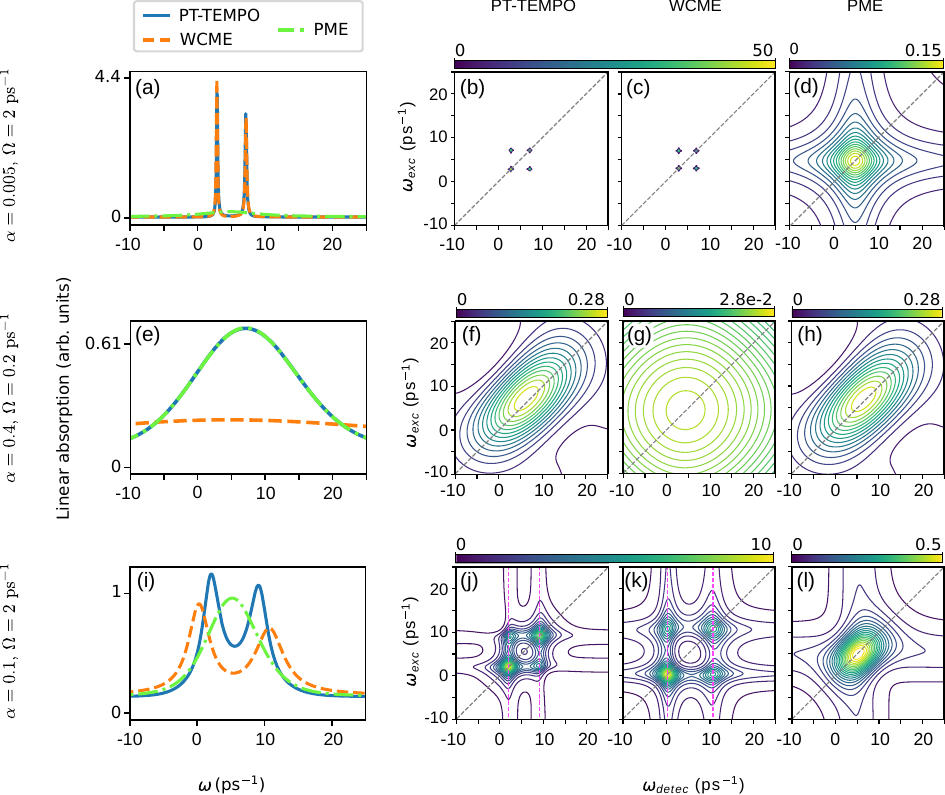}
\caption{Comparison of linear absorption and 2D spectra in (a-d) a weak system-bath coupling regime, (e-h) a strong-coupling (polaronic) regime and (i-l) an intermediate-coupling regime, calculated using PT-TEMPO (blue solid in linear spectra), a weak-coupling master equation (WCME; orange dashed in in linear spectra) or a polaron master equation (PME; green dot-dashed in linear spectra). All spectra were calculated at $T = 13$~ps$^{-1}$ (100 K), using dipole operator $\hat{V}_{2}$. For (j) and (k), the vertical purple dashed lines mark the diagonal peak maxima.}
\label{fig3}
\end{figure*}

\section{Master equation comparison}
\label{sec:MECompare}
To test the validity of our method, we compare spectra obtained with PT-TEMPO against a Markovian weak-coupling master equation (WCME) and polaron master equation (PME). To do this, we first present the results in two different regimes: a weak system-bath coupling regime ($\alpha=0.005$) where the WCME is expected to give accurate results, and a polaron regime with strong system-bath coupling ($\alpha=0.4$) and reduced electronic coupling ($\Omega=0.2$~ps$^{-1}$), suitable for the PME.
As with PT-TEMPO, the master equation spectra were simulated with OQuPy~\cite{oqupy}, which has a Lindblad master equation solver implemented that works in tandem with the multi-time correlations module.

\subsection{Weak-coupling master equation}
Starting with the WCME, we obtain the following master equation in Lindblad form using standard techniques~\cite{breuer:2002}:
\begin{multline}
    \dv{t}\rho_S = -i\bqty{\hat{H}^\prime_S, \rho_S} \\
    + \sum_{n}^2 \gamma_n \pqty{\hat{L}_n \rho_S \hat{L}_n^\dagger -\frac{1}{2}\Bqty{\hat{L}_n^\dagger\hat{L}_n, \rho_S}},
    \label{me}
\end{multline}
where $\hat{L}_1 = \ketbra{-}{+}$ and $\hat{L}_2 = \hat{L}_1^\dagger$. The rates $\gamma_n$ are given by $\gamma_1 = 2\pi J\pqty{2\Omega}\pqty{N\pqty{2\Omega} +1}$ and $\gamma_2 =2\pi J\pqty{2\Omega}N\pqty{2\Omega} $, where $N(\omega)$ is the Bose-Einstein occupation number and $J(\omega)$ is the spectral density in Eq.~\eqref{sd}. The electronic energies are furthermore modified by a Lamb shift so that
\begin{equation}
    \hat{H}^\prime_S = \pqty{E_- + S(-2\Omega)}\dyad{-}{-} +\pqty{E_+ + S(2\Omega}) \dyad{+}{+},
\end{equation}
in the eigenbasis $\ket{\pm}$ with $E_\pm = \epsilon+\lambda \pm \Omega$, and
\begin{equation}
    S(\nu) = \mathcal{P} \int^\infty_{-\infty} \frac{J(\omega)\pqty{N(\omega)+1} +J(-\omega)N(-\omega)}{\nu - \omega}.
    \label{lamb}
\end{equation}
Here $\mathcal{P}$ denotes the Cauchy principal value and we define $J(\omega) = 0$ for $\omega < 0$. 

We note that
$S(+2\Omega)$ corresponds to a positive shift to $E_+$, while $S(-2\Omega)$ corresponds to a negative shift to $E_-$. The total energy splitting between the excited eigenstates therefore becomes:
\begin{flalign}
    \delta E &= 2\Omega + S(+2\Omega) - S(-2\Omega)\\
     &= 2\Omega\pqty{1+ 2\int_0^\infty \dd \omega \frac{J(\omega)\coth\pqty{\omega/(2k_B T)}}{\pqty{2\Omega}^2 - \omega^2}}.
    \label{splitting}
\end{flalign}

As illustrated in Figs.~\ref{fig3}(a-c), the PT-TEMPO simulations of linear absorption and 2D spectra agree well with the WCME in the weak-coupling regime. In the 2D spectra, the two peaks along the diagonal of the spectra reflect the transition frequencies of the model, which coincide with the peak frequencies in the linear absorption spectrum. For $\alpha=0.005$, the Lamb shift, Eq.~\eqref{lamb}, is small compared to $\epsilon$ and $\Omega$, such that the splitting between peaks is close to the bare splitting $2\Omega$. The 2D spectra furthermore contain two cross-peaks that correlate the transition frequencies and are a signature of the coupling between the electronic excited states.

For some problems with weakly-structured environments, {\it i.e.} those with non-constant $J(\omega)$, one can improve on the Lindblad master equation by restoring non-secular terms from Redfield theory~\cite{Eastham2016Bath,Hartmann2020Accuracy}.
For our particular choice of dipole operators and initial conditions, the non-secular terms in the master equation do not affect the outcome. To show this, consider the individual components of the reduced system density matrix
\begin{equation}
    \rho_S = \left( \begin{array}{ccc} \rho_{00} & \rho_{0-} & \rho_{0+} \\ \rho_{-0} & \rho_{--} & \rho_{-+} \\ \rho_{+0} & \rho_{+-} & \rho_{++}\end{array} \right),
    \label{rho_c}
\end{equation}
expressed in terms of system eigenstates. The non-secular terms in the master equation would have the form:
\begin{equation}
    \label{eq:nonsecular}
    \sum_{n}^2 \gamma_n \pqty{\hat{L}_n \rho_S \hat{L}_n + \hat{L}^\dagger_n \rho_S \hat{L}^\dagger_n},
\end{equation}
where $\hat{L}_1 = \ketbra{-}{+}$, $\hat{L}_2 = \hat{L}_1^\dagger$ and $\gamma_n$ are the transition rates stated in the main text. Note here that $\hat{L}_n\hat{L}_n = \hat{L}_n^\dagger \hat{L}_n^\dagger = 0$, and terms involving these are therefore not included in Eq.~\eqref{eq:nonsecular}. For the individual components of $\rho_S$ in Eq.~\eqref{rho_c}, we find that only $\rho_{-+}$ and $\rho_{+-}$ would be modified by these non-secular terms. The dipole operator $\hat{V}_2 = \ketbra{0}{2} + \mathrm{H.c.}$, on the other hand, 
only creates $\rho_{0\pm}$ and $\rho_{\pm 0}$ components when applied to the initial state $\ketbra{0}{0}$. Thus, for the absorption spectra presented in this work, there would be no difference between secular and non-secular master equations. For the 2D spectra, $\rho_{+-}$ and $\rho_{-+}$ components would be created after application of the second dipole operator at time $\tau_2$, Eq.~\eqref{R}. Since in this work we are only considering multi-time correlations for which $\tau_2 = \tau_3$ (the third operator is applied immediately after the second), there would be no difference for the 2D spectra presented here either.

\subsection{Polaron master equation} \label{section:pme}

To derive a master equation for our three-level system, Eq.~\eqref{ham}, in the strong-coupling regime, we first apply a polaron transformation to the Hamiltonian~\cite{holstein:1959, jang:2008, pollock:2013, nazir:2016}:
\begin{equation}
    \Tilde{H} = e^G \hat{H} e^{-G}; \,\,\,\,\,\,\, G = \sum_k \frac{g_k}{\omega_k}\big(b^\dagger_k - b_k\big)\big(\vert 1 \rangle\langle 1 \vert - \vert 2 \rangle\langle 2 \vert\big),
    \label{polarontransf}
\end{equation}
to obtain:
\begin{flalign}
    \Tilde{H}_S &= \epsilon \pqty{\ketbra{1}{1} + \ketbra{2}{2}} + \Omega^\prime\pqty{\ketbra{1}{2} + \mathrm{H.c.}} \nonumber \\
    \Tilde{H}_B &= \hat{H}_B  \nonumber \\
    \Tilde{H}_I &= \Omega\mathcal{D}_c\vert 1 \rangle \langle 2 \vert + \mathrm{H.c.}  \label{htilde}.
\end{flalign}
In the polaron frame, the interaction Hamiltonian $\hat{H}_I$ contains the coupling operator $\mathcal{D}_c$, which depends on the displacement operator
\begin{equation}
    \mathcal{D} = \exp{\sum_k \frac{2g_k}{\omega_k}(b^\dagger_k - b_k)},
\end{equation}
such that $\mathcal{D}_c$ is given by:
\begin{equation}
    \mathcal{D}_c = \mathcal{D} - \langle \mathcal{D} \rangle.
    \label{displ}
\end{equation}
Here, the expectation value of $\mathcal{D}$
\begin{equation}
    \langle \mathcal{D} \rangle = \exp{-2\sum_k \frac{g_k^2}{\omega_k^2}\coth{\Big(\frac{\beta\omega_k}{2}\Big)}}
\end{equation}
has been subtracted from $\mathcal{D}$  and is instead added to $\hat{H}_S$, which now contains the renormalized electronic coupling:
\begin{equation}
    \Omega^\prime = \Omega\langle \mathcal{D} \rangle.
\end{equation}
Note here that for our choice of an Ohmic spectral density in Eq.~\eqref{sd}, $\expval{\mathcal{D}}$ (and therefore $\Omega^\prime$) are infinitesimally small. 
Following the same standard techniques as for the WCME, we can now perturbatively expand $\Tilde{H}_I$ to find the polaron master equation:
\begin{multline}
    \dv{t}\tilde{\rho}_S = -i\bqty{\tilde{H}^\prime_S, \tilde{\rho}_S} \\
    + \sum_{n}^3 \Lambda_n \pqty{\hat{L}_n \tilde{\rho}_S \hat{L}_n^\dagger -\frac{1}{2}\Bqty{\hat{L}_n^\dagger\hat{L}_n, \tilde{\rho}_S}},
    \label{pme}
\end{multline}
with Lindblad operators $\hat{L}_1 = \ketbra{-}{+} = \hat{L}_2^\dagger$ and $\hat{L}_3 = \ketbra{-}{-} - \ketbra{+}{+}$. 
The transition rates $\Lambda_n$ depend on the four bath correlation functions $\langle \mathcal{D}_c(t)\mathcal{D}_c(0)\rangle, \langle \mathcal{D}_c^\dagger(t)\mathcal{D}_c^\dagger(0)\rangle, \langle \mathcal{D}_c^\dagger(t)\mathcal{D}_c(0)\rangle$ and $\langle \mathcal{D}_c(t)\mathcal{D}_c^\dagger(0)\rangle$. For example:
\begin{equation}
    \langle \mathcal{D}_c(t)\mathcal{D}_c^\dagger(0)\rangle = \langle e^{i\hat{P}(t)} e^{-i\hat{P}(0)}\rangle - \expval{\mathcal{D}}^2,
    \label{dipcor}
\end{equation}
where $\hat{P}(t) = -i \sum_k \frac{2g_k}{\omega_k}\pqty{b_k^\dagger(t) - b_k(t)}$.
To solve Eq.~\eqref{dipcor}, we can add the exponents together using the Baker-Campbell-Hausdorff formula:
\begin{equation}
    \langle \mathcal{D}_c(t)\mathcal{D}_c^\dagger(0)\rangle = \expval{e^{i\hat{P}(t) -i\hat{P}(0) + \frac{1}{2} \bqty{i\hat{P}(t), -i \hat{P}(0)}}} - \expval{\mathcal{D}}^2.
    \label{dipcor2}
\end{equation}
Then, since the problem is Gaussian in the bosonic operators, we can shift the expectation to the exponent. In our case $\expval{\hat{P}}=0$, so only the second order term in $\hat{P}$ survives:
\begin{equation}
    \expval{e^{\hat{P}}} = \exp \bqty{ \frac{1}{2}\expval{\hat{P}^2}      }.
\end{equation}
This leaves us with the following expression:
\begin{equation}
    \langle \mathcal{D}_c(t)\mathcal{D}_c^\dagger(0)\rangle = \exp \bqty{\frac{1}{2}\expval{\hat{P}(0)^2} + \expval{\hat{P}(t)\hat{P}(0)}} - \expval{\mathcal{D}}^2,
\end{equation}
which we can further simplify by defining the phonon propagator
\begin{equation}
    \phi(t) = \int^\infty_0 \dd \omega \frac{J(\omega)}{\omega^2}\bigg(\cos (\omega t) \coth \Big(\frac{\omega}{2k_B T} \Big)- i \sin (\omega t)      \bigg).
    \label{phononprop}
\end{equation}
Note here that for an Ohmic spectral density, the integrand of $\phi(0)$ diverges as $\omega \rightarrow 0$, meaning that $e^{-4\phi(0)} = 0$. Applying this limit to all four bath correlation functions leaves us with:
\begin{flalign}
    &\expval{\mathcal{D}_c(t)\mathcal{D}_c^\dagger(0)} = \expval{\mathcal{D}_c^\dagger(t)\mathcal{D}_c(0)}= e^{4 \pqty{\phi(t)-\phi(0)}}; \nonumber \\
    &\expval{\mathcal{D}_c^\dagger(t)\mathcal{D}_c^\dagger(0)} = \expval{\mathcal{D}_c(t)\mathcal{D}_c(0)} = 0.
\end{flalign}
Finally, taking the time integral over these correlations and inserting the coupling $\Omega$ gives us the transition rates $\Lambda_n$ in the master equation in Eq.~\eqref{pme}:
\begin{flalign}
    \Lambda_1 &= \Lambda_2 = \int_0^\infty\dd t\, \Omega^2 e^{4 \pqty{\phi(t)-\phi(0)}}, \nonumber \\
    \Lambda_3 &= 2\Lambda_1,
    \label{polaronrates}
\end{flalign}
which can be solved numerically. As for the WCME, the imaginary parts of these integrals correspond to the Lamb shift, and are incorporated into $\tilde{H}^\prime_S$. For the PME however, we find that the Lamb shift is about two orders of magnitude smaller than system frequency scales, and thus has a limited impact on the peak locations in Fig.~\ref{fig3}.

As shown in Fig.~\ref{fig3}(e-g), the PME results agree well with PT-TEMPO in the polaron regime ($\alpha = 0.4$, $\Omega = 0.2$~ps$^{-1}$). Unlike the WCME spectrum (Fig.~\ref{fig3}(g)) in which the peak is homogeneously broadened in all directions, the PT-TEMPO and PME results capture correlations between excitations at $t_1$ and $t_3$, leading to a inhomogeneously broadened peak stretched along the diagonal axis. For the PME, the transition rates $\Lambda_n$ in Eq.~\eqref{polaronrates} are negligibly small. Rather, the peak broadening in Fig.~\ref{fig3}(h) comes from the polaron transformation in Eq.~\eqref{polarontransf}, which encodes optical dephasing between the ground and excited states.

Because $\tilde{H}_I$ is proportional to $\Omega$ (see Eq.~\eqref{htilde}), larger values of $\Omega$ would not give correct results. This is illustrated in Fig.~\ref{fig3}(a)(d), which shows the 2D and linear spectrum obtained with the PME in the weak-coupling regime ($\alpha = 0.005$, $\Omega=2.0$~ps$^{-1}$). Since $\Omega^\prime$ is infinitesimally small, only a single peak (and therefore no cross-peaks) can be resolved. Additionally, because the rates in Eq.~\eqref{polaronrates} are proportional to $\Omega^2$, this peak is strongly broadened compared to the PT-TEMPO and WCME solutions.

\begin{figure}
\includegraphics[width=8cm]{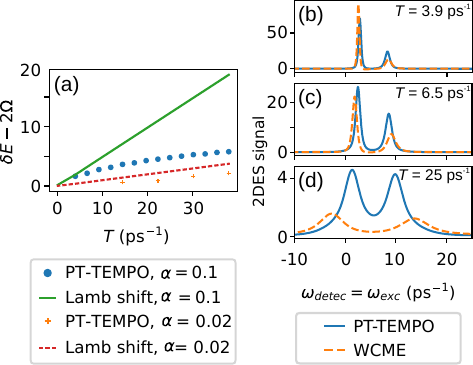}
\caption{(a) Splitting between diagonal peaks in 2D spectra as a function of temperature, as predicted by PT-TEMPO and the WCME (Lamb shift). Data is shown for system-bath couplings $\alpha=0.1$ (green solid, blue dots) and $\alpha=0.02$ (red dashes and orange cross markers). Note that due to increasing computational resources with decreasing $T$, data points for PT-TEMPO below $3.9$~ps$^{-1}$ (30~K) are not shown. (b-d) Diagonal slices through the 2D spectra for $\alpha=0.1$, corresponding to the data points $T=3.9$~ps$^{-1}$ (30~K), $T=6.5$~ps$^{-1}$ (50~K) and $T=25$~ps$^{-1}$ (190~K) in panel (a). For all: $\Omega = 2$~ps$^{-1}$.}
\label{fig4}
\end{figure}

\section{Differences in the intermediate coupling regime}
\label{sec:Differences}
We will now shift our focus to the intermediate coupling regime found in photosynthetic complexes, where neither the weak-coupling nor polaron master equation are expected to give accurate results. For this purpose, we will set  $\Omega = 2$~ps$^{-1}$ and $\alpha = 0.1$~\cite{ishizaki:2010, fassioli:2014, jumper:2018}. As in the weak-coupling regime, the 2D spectrum obtained with PT-TEMPO for these parameters, Fig.~\ref{fig3}(i), contains two diagonal peaks and cross-peaks, although the peaks are more broadened due to the stronger system-bath coupling. Firstly, we observe that the PME spectra, Fig.~\ref{fig3}(i)(l), cannot replicate the double peak structure found in the numerically exact result, because the electronic coupling $\Omega$ is renormalized to zero (see Section~\ref{section:pme}). The WCME spectra, Fig.~\ref{fig3}(i)(k), on the other hand do accurately reflect the presence of all peaks. For the chosen parameters however, the Lamb shift in the WCME spectra, Eq.~\eqref{splitting}, overestimates the splitting between the peaks compared to the PT-TEMPO results. This difference is illustrated in more detail in Fig.~\ref{fig4}(a): for both PT-TEMPO and the WCME, the relative splitting between peaks ($\delta E - 2 \Omega$, Eq.~\eqref{splitting}) increases with temperature. For the WCME, this increase is linear for large temperatures ($\frac{\omega_c}{2k_B T} \ll 1$). Although the splitting predicted by PT-TEMPO agrees well with the Lamb shift below $T\approx 4$~ps$^{-1}$, an increasing discrepancy arises when the temperature is increased. Since the Lamb shift is directly proportional to $\alpha$ for our choice of spectral density, Eq.~\eqref{sd}, this discrepancy decreases when $\alpha$ is decreased, depicted by the red dashes and orange cross markers in Fig.~\ref{fig4}(a) for $\alpha=0.02$. 

While the WCME gives more accurate predictions for peak splitting in the intermediate coupling regime as temperature is decreased, its prediction for peak height breaks down. As shown in Figs.~\ref{fig4}(b-d), the ratio between diagonal peak amplitudes increases with decreasing temperature. While this is also the case for the numerically exact results, the WCME predicts a sharper and higher amplitude peak for the $\ket{-}$ transition at sufficiently low $T$ (Fig.~\ref{fig4}(b)), and similarly a broader peak for $\ket{+}$ transition. 
From the WCME in Eq.~\eqref{me}, we can observe that the transition rate $\gamma_1$ is always finite, while $\gamma_2$ tends to zero as temperature decreases. As a result of this and the absence of a pure dephasing rate, the optical response predicted by the WCME decays more slowly at low temperatures compared to PT-TEMPO, leading to a more pronounced peak corresponding to the $\ket{-}$ state~\cite{chin:2013}. As temperature increases on the other hand, Fig.~\ref{fig4}(c), $\gamma_1$ and $\gamma_2$ become more comparable in magnitude, decreasing the relative difference between diagonal peak amplitudes.

\section{Conclusions}
\label{sec:conclusions}
In this work, we have presented a tensor network method for the efficient calculation of multi-time correlation functions, based on a process tensor framework. We have employed our technique to simulate linear absorption and 2D spectra in three different system-bath coupling regimes, comparing the results to a weak-coupling and polaron master equation. Here we observed that in an intermediate coupling regime, both master equations break down in the following ways: for an Ohmic spectral density, a polaron master equation fails to resolve the two energy transitions probed in our model. A weak-coupling master equation on the one hand overestimates peak splitting (given by the Lamb shift) at high temperatures. At low temperatures on the other hand, it overestimates the decay time of the optical response, leading to discrepancies in peak amplitude ratios compared to our numerically exact results.

We furthermore note that several PT-MPO methods were recently developed that exploit time translational invariance to improve the scaling of the algorithm with the number of time steps~\cite{cygorek:2023update, link:2023}. Employing such algorithms could further reduce the computational effort of calculating multi-time correlation functions with process tensors. 

Finally, we hope that beyond light-harvesting, the methods employed in this work may aid in the study of other non-Markovian open quantum systems, such as coherent dynamics in semiconductor quantum dots and the polaronic wave functions of excitons in perovskite materials~\cite{cassette:2016, smallwood:2018, collini:2019, tao2022dynamic}.

\begin{acknowledgments}
    For insightful comments on an earlier version of this paper, we would like to thank Martin Plenio, Susana Huelga, Nicola Lorenzoni and Jaemin Lim. R.\ d.\ W. acknowledges support from EPSRC (EP/W524505/1).  B.\ W.\ L.\ and J.\ K.\ acknowledge support from EPSRC (EP/T014032/1). A.\ W.\ C.\ wishes to acknowledge support from ANR Project ACCEPT (Grant No. ANR-19-CE24-0028).
\end{acknowledgments}

\bibliography{references}

\clearpage
\newpage
\renewcommand{\theequation}{A\arabic{equation}}
\renewcommand{\thefigure}{A\arabic{figure}}
\setcounter{section}{0}
\setcounter{equation}{0}
\setcounter{figure}{0}
\setcounter{page}{1}

\section*{Appendix A: Computational resources} 
\label{app:comp}

For this work, the required PT-MPOs were constructed using the open source package OQuPy~\cite{oqupy, oqupaper}. For the Ohmic spectral density used in this work, $J(\omega) = 2 \alpha \omega e^{-\frac{\omega}{\omega_c}}$ ($\alpha = 0.1$, $\omega_c = 3.04$~ps$^{-1}$), it took 8.7~mins to construct a PT (150 time steps) at a temperature $T = 13.09$~ps$^{-1}$ on a single core of an Intel i5 (8th Gen) processor. Computing a single four-time correlation function as a function of $\tau_4$ ({\it e.g.}  $R(\tau_4; \tau_1=\tau_2=\tau_3=0)$ took 101~s on a single CPU core. This computation time scales linearly with the number of additional time steps when one of the earlier time arguments ($\tau_{1,2,3}$) is varied. For example, the 2D spectra in Fig.~\ref{fig2} were computed over 50 time steps each in $\tau_1$ and $\tau_4$. Taking into account that it is composed of four correlation functions (Eq.~\eqref{R}), the total 2D spectrum in Fig.~\ref{fig2}(a) required 5.6 core hours with a pre-computed PT.

For a set number of time steps, constructing the PT at lower temperatures reduces the computation time. However, since multi-time correlations generally take longer to decay to zero with decreasing temperature, a longer PT has to be constructed to capture the full signal. For example, the PT constructed for Fig.~\ref{fig4}(a) and $T=3.9$~ps$^{-1}$ was 300 time steps in length and took 12~mins to compute on a single CPU core.

\section*{Appendix B: Convergence of multi-time correlations}
\label{app:conv}
\begin{figure}
\includegraphics[width=8cm]{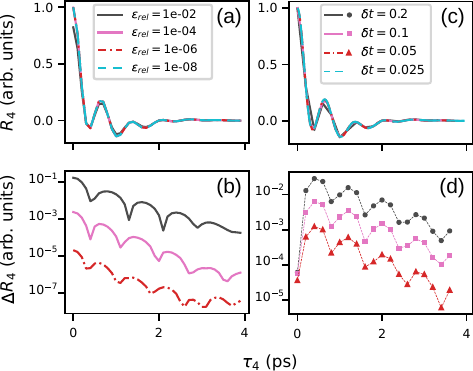}
\caption{Numerical convergence of the four-time correlation function $R_4(\tau_4)$ (Eq.~\eqref{R40}) with respect to (a-b) the truncation error $\epsilon_{rel}$ for a constant time step $\delta t = 0.1$~ps and (c-d) with respect to $\delta t$ and $\epsilon_{rel} = 10^{-6}$. Figures (b) and (c) show the absolute difference between $R_4$ calculated with the most precise convergence parameter ((b) $\epsilon_{rel} = 10^{-8}$, (c) $\delta t=0.025$~ps) and with each lower precision. }
\label{s1}
\end{figure}

The PT-TEMPO algorithm relies on three computational parameters: the time step $\delta t$, the maximum memory length $\Delta K_{max}$ and the maximal relative error in the singular value cutoff $\epsilon_{rel}$ \cite{oqupaper}. The product $\delta t \Delta K_{max}$ corresponds to the maximal memory time of the bath captured by the computations. Therefore, the value of $\delta t \Delta K_{max}$ should be larger than the time it takes for the bath autocorrelation function to decay to zero:
\begin{equation}
    C(\tau) = \int_0^\infty \dd \omega J(\omega) \bqty{\cos (\omega \tau) \coth \pqty{\frac{\omega}{2 k_B T}}  - i \sin (\omega \tau)}.
\end{equation}
For an Ohmic spectral density (Eq.~\eqref{sd}) with $\alpha = 0.1$ and $\omega_c = 3.04$~ps$^{-1}$, we find that $C(\tau)$ decays to a value $10^{-3}$ times smaller than  its maximum when $\tau = 10.3$~ps. In this work, we have set $\epsilon_{rel} = 10^{-6}$, $\Delta K_{max} = 1000$ and $\delta t = 0.05-0.1$~ps, which places the cutoff well beyond the memory time of the bath. To justify our choice of $\delta t$ and $\epsilon_{rel}$, we test for numerical convergence by computing the four-time correlation function  $R_4(\tau_4; \tau_1=\tau_2=\tau_3=0)$ in Eq.~\eqref{R}:
\begin{equation}
    R_4(\tau_4) = \Tr \big[\hat{V}_2(\tau_4)\hat{V}_2(0) \hat{V}_2(0)\hat{V}_2(0) \rho_0 \big].
    \label{R40}
\end{equation}
Fig.~\ref{s1}(a) shows $R_4(\tau_4)$ for a range of $\epsilon_{rel}$ values and a constant time step $\delta t = 0.1$~ps. To better illustrate the convergence, the absolute difference $\Delta R_4$ between the curve with the highest precision ($\epsilon_{rel}= 10^{-8}$) and each subsequent lower precision curve is plotted in Fig.~\ref{s1}(b). Similarly, Fig.~\ref{s1}(d) shows $\Delta R_4$ for different $\delta t$, taking $\delta t = 0.025$~ps as the baseline. 

\end{document}